# Implementación de un Filtro de Polarización basado en la Descomposición del Valor Singular (SVD)

**Rubén Luque**[1,2]; **Nuri Hurtado**[1] y **Asdrúbal Ovalles**[2]

[1]*Laboratorio de Física Teórica de Sólidos. Escuela de Física. CEFITEC. Universidad Central de Venezuela, Caracas, Venezuela.*

[2]*Centro de Procesamiento de Datos Geofísicos. PDVSA-INTEVEP.*

## Resumen

En este trabajo hemos introducido una técnica adaptativa de filtrado, que permite resaltar los eventos sísmicos selectivamente de acuerdo a su polarización. Esta técnica consiste en la implementación de un filtro de polarización basado en la Descomposición del Valor Singular (SVD). El SVD tiene una gran variedad de aplicaciones en el procesamiento de señales e imágenes, en este caso, un registro sísmico es considerado una imagen del subsuelo. Con esta técnica, se logró la reconstrucción de la imagen sísmica generada a partir de las primeras dos autoimágenes, reproducir los atributos de polarización (rectilinialidad y planaridad) del movimiento de las partículas, suprimir el ruido aleatorio de alta frecuencia, mejorar la relación señal-ruido y de esta forma se obtuvo una imagen clara y coherente de los eventos sísmicos asociados a las energías primarias de reflexión.

**Palabras clave:** *SVD, Filtros de Polarización, Autoimágenes, Factores de Rectilinialidad y Planaridad, Ruido Sísmico, Relación Señal-Ruido.*

## Abstract

In this job we have introduced a filtered adaptive technique that allows highlighting selective seismic events according to its polarization. This technique consists in the implementation of a polarizing filter based on the Single Value Decomposition (SVD). The SVD have a great variety of application in the processing of signals and images, in this case a seismic reading is considered an image of the underground. With this technique, was achieved to make a reconstruction of the seismic image generated after the two first autoimages reproduce the polarization attributes (rectilinearity and planarity) of the moving particles, suppress the high frequency random sound, improve the signal-noise ratio and this way we were able to obtain a clear and coherent image of the associated seismic events to the primary energies of reflection.

# Introducción

En la exploración geofísica de hidrocarburos, uno de los métodos de mayor aplicabilidad es el método sísmico de reflexión. Este, es utilizado para delinear la geología del subsuelo (Yilmaz, 1987), y consiste en registrar las vibraciones del suelo en superficie, cuando son producidas por una fuente artificial (explosivos, camiones vibradores, caída en peso, etc.). En la etapa del procesamiento de los datos sísmicos registrados, para diferentes estudios, se han aplicado diversas técnicas de filtrado (Yilmaz, 1987; De Meersman y Kendall, 2005; De Meersman y Ansorger, 2007), con la finalidad de suprimir las bandas de frecuencias de energías no útiles, que son conocidas como "ruido sísmico".

Un filtro no convencional, pero que en la última década ha sido considerado una importante técnica en el procesamiento de señales en la exploración sísmica, basada en la reconstrucción de la imagen, la supresión del ruido y acrecentar la relación señal-ruido (Bekara y Van der Baan, 2007), es la técnica de la Descomposición del Valor Singular (SVD). La SVD consiste en una factorización significativa de una matriz rectangular, real o compleja, que se basa en el estudio de las componentes de los ejes principales, de un número de parámetros o datos que se desean comprimir o simplificar (Forsythe, 1977).

El SVD, tiene una variedad de aplicaciones en el procesamiento de señales (Forsythe, 1977). Aunque, esta técnica está enfocada como una poderosa matriz de descomposición (Klema y Lamb, 1980; Ursin y Zheng, 1985), también ha sido mostrada desde un punto de vista equivalente en la aplicación del procesamiento de imágenes (Andrews y Hunt, 1977), como un algoritmo eficaz que permite resaltar eventos principales (asociado a energías) y a su vez la reconstrucción de las imágenes, donde la descomposición de la matriz de datos de entrada es descrita en autoimágenes (Freire y Ulrych, 1988).

En el presente trabajo, se desarrolla una metodología con la finalidad de atenuar uno de los ruidos sísmicos que más afecta la exploración sísmica, como lo es el ruido aleatorio de alta frecuencia. Este tipo de ruido, enmascara e inclusive, en algunos casos se solapa (en frecuencia) con el dato sísmico (reflexiones de interés). La metodología en cuestión, corresponde a la implementación del filtro de polarización SVD, el cual se basa en técnicas del algebra lineal, y es aplicado a datos símicos multicomponente.

# Metodología

La metodología desarrollada en este trabajo, está basada en conformar una matriz de datos que llamamos Matriz X, a partir de la tripleta de matrices correspondientes a las componentes de los registros sísmicos generados: vertical, Z, radial, R, y transversal, T. Donde las dimensiones MxN en cada matriz, representan el número de muestras (M) por el número de geófonos (N). El algoritmo calculará los valores singulares de la Matriz X y las dos primeras autoimágenes ($E_1$ y $E_2$), asociadas a la energía de la señal polarizada elípticamente, la cual se encuentra principalmente sobre los dos primeros ejes principales de esta señal.

Asumiendo que el ruido es aleatoriamente polarizado, Jurkevics (1988), establece que las funciones adecuadas en conjunto con la ponderación de las dos primeras autoimágenes para generar mejoras en la relación señal-ruido, son los factores de rectilinialidad $R_1$ y $R_2$, correspondientes al primer y segundo eje principal, así como la planaridad, P:

$$R_1 = 1 - \left(\frac{\sigma_3^2}{\sigma_1^2}\right); \quad R_2 = 1 - \left(\frac{\sigma_3^2}{\sigma_2^2}\right);$$

$$P = 1 - (2\sigma_3^2/(\sigma_1^2 + \sigma_2^2))$$

donde $\sigma_i^2$ ($\sigma_i$, i-ésimo Valor Singular de la matriz $X$, y $\sigma_1 \geq \sigma_2 \geq \sigma_3 \ldots \geq \sigma_M$) es la energía de los datos de las componentes principales a lo largo de $v_i$ (i-ésimo autovector de $X^T X$).

El conjunto de los factores calculados y la ponderación de las dos primeras autoimágenes, generan la ecuación del operador del filtro basado en la técnica SVD, mediante el cual se obtiene la señal filtrada de los datos sísmicos multicomponente, que esta expresada por:

$$F = \sum_i^2 (u_i \sigma_i v_i^T R_i) P = (E_1 R_1 + E_2 R_2) P, \quad (a)$$

donde $F$ es la señal filtrada y $u_i$ es el i-ésimo autovector de $X^T X$. Esta ecuación representa el caso más general de la polarización elíptica, donde la polarización lineal representa un caso particular ($F \cong E_1(R_1)^2$).

# Resultados

Con la implementación del filtro de polarización basado en la Descomposición del Valor Singular (SVD), el cual, es un filtro adaptativo orientado a atenuar el ruido sísmico, hemos logrado preservar eventos de polarización lineal, realzar eventos sísmicos (reflexiones de interés) presentes en los registros y así como acrecentar la relación señal ruido. Básicamente, el algoritmo trabaja en una ventana de tiempo suficientemente larga (80mseg), como para abarcar la longitud de la ondícula correspondiente al evento asociado a la fase con polarización lineal.

Comparando las imágenes b) y c) de la figura 1, podemos observar que el filtro pasa-banda (PB) suprime parte del ruido aleatorio (ruido en la banda de frecuencia de 20Hz a 80Hz) presente en el registro sin filtrar (fig. 1a). Así mismo se puede observar que la energía asociada al ruido no es atenuada por completo, sino que está superpuesta sobre los eventos sísmicos de interés, ocasionando la degradación de la calidad de los datos, a través de los cuales se describe la geología del subsuelo. Con la utilización del filtro PB se preservan los eventos sísmicos, pero con una forma suavizada de la ondícula, ya que éste no es capaz de reconocer si en esa banda de frecuencias está contenida señal útil o no.

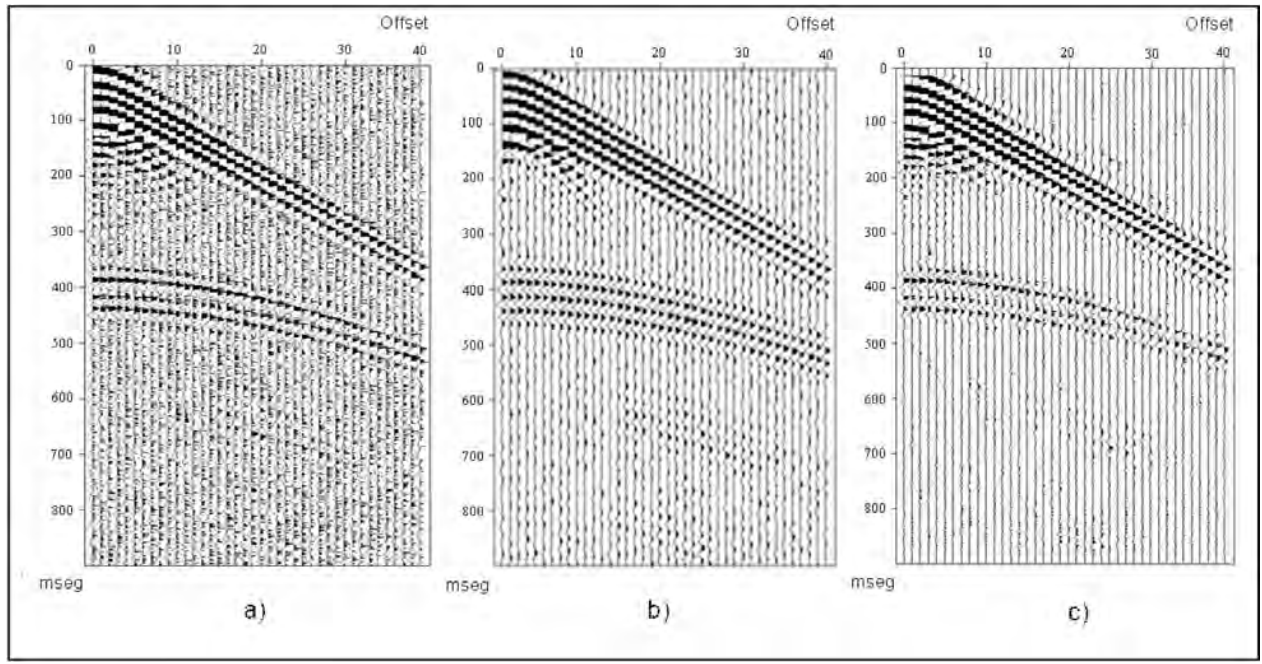

**Figura 1.** Componente Vertical "Z" de los Registros Sísmicos Sintéticos: a) Registro sin filtrar, con ruido aleatorio en la banda de frecuencia de 20Hz a 80Hz, b) Registro filtrado a través de un filtro Pasa-Banda (PB) en un rango de frecuencias de 12-16Hz a 40-60Hz y c) Registro filtrado utilizando la técnica de la Descomposición del Valor Singular (SVD)

El filtro de polarización basado en la SVD, que cubre todos los rangos de frecuencias, nos permitió encontrar resultados más satisfactorios (fig. 1c) que los encontrados con el filtro PB (fig. 1b), cuyo rango de frecuencias es de 12-16Hz a 40-60Hz. Un aspecto importante y ventajoso del filtro SVD, es que está enfocado en la proyección del movimiento de las partículas hacia los ejes principales, donde está contenida la mayor parte de la información de energía, que está asociada a los eventos sísmicos de reflexiones primarias. Así como también se logran reproducir los atributos de polarización de los datos sísmicos multicomponente y la reconstrucción de la imagen sísmica.

Otra ventaja de este filtro es que se lograron preservar los eventos sísmicos asociados a polarizaciones lineales, sin que ocurriese el suavizado de ondículas, con lo cual se evitó la generación de eventos inexistentes. De igual forma se logró realzar y aumentar la coherencia lateral de los eventos sísmicos, así como acrecentar la relación señal-ruido.

## Conclusiones

Implementamos, sobre datos sísmicos multicomponente que fueron generados en el programa de modelado elástico "e3d_anel21", un filtro de polarización basado en la Descomposición del Valor Singular (SVD), donde el filtro suprime exitosamente la mayoría de la energía del ruido aleatorio de alta frecuencia.

El filtro de polarización basado en la técnica SVD, demostró ser una herramienta más eficaz en comparación al filtro convencional pasa-banda, debido a que realza y aumenta la coherencia lateral de los eventos sísmicos, reproduce los atributos de polarización de los datos sísmicos multicomponente, así como la amplitud y fase de la señal, proporcionando información coherente para la reconstrucción de la imagen sísmica y de esta forma describir las características y propiedades del subsuelo. Gracias a esta técnica se logró acrecentar la relación señal-ruido.

## Agradecimientos

## Referencias

Andrews H. C.; Hunt B. R. (1977). "*Digital image restoration*". Prentice-Hall, Signal Processing Series.

Bekara, M.; Van der Baan, M. (2007). "*Local Singular Value Decomposition for Signal Enhancement of Seismic Data*". Geophysics, Vol.72, 59-65.

De Meersman, K.; Ansorger, C. (2007). "*Ground Roll Removal and Signal Preservation by Cascading SVD Polarization Filters with Localized Fk-Filters*". CSPG CSEG Convention. 450.

De Meersman,; Kendall, R. (2005). *A complex SVD-polarization filter for ground roll attenuation on multi-component data*. CSEG Convention.

Forsythe G. E.; Malcolm M. A.; Moler, C. B. (1977). "*Computer Methods for Mathematical Computations*". Cap. 9.

Freire S. L. M.; Ulrych, T. J. (1988). "*Application of singular value decomposition to vertical seismic profiling*", Geophysics, Vol.53, 778-785.

Jurkevics, A. (1988). "*Polarization Analysis of three-component array data*". Bull. Seis. Soc Am., 78, 1725-1743.

Klema, V. C.; Lamb, A. J. (1980). "*The Singular Value Decomposition: its computation and some applications*". Inst. Electr. and Electron. Eng., Trans. Automatic Control. AC-25, 164-176.

Ursin, B.; Zheng, Y. (1985). "*Identification of Seismic Reflection using Singular Value Decomposition*". Geophys. Prosp., 33, 773-779.

Yilmaz, OZ., (1987). "*Seismic Data Analysis*". Society of Exploration Geophysicists, Volumen I.